\documentclass[a4paper,11pt]{article}

\usepackage{lineno}
\usepackage{slashed}
\usepackage{adjustbox}
\usepackage{physics}
\usepackage{tikz}
\usetikzlibrary{decorations.markings}
\usepackage{amsmath,amssymb}
\usepackage{float}
\usepackage{graphicx}
\usepackage{menukeys}
\usepackage{slashed}
\usepackage{bbold}
\usepackage{xr}
\usetikzlibrary{arrows,shapes,patterns,calc}
\usepackage[utf8]{inputenc}
\usepackage{simpler-wick}
\usepackage{comment}
\usepackage{amssymb}
\usepackage{amsmath}
\usepackage{amsfonts}
\usepackage{array}
\usepackage{multirow}
\usepackage{psfrag}
\usepackage{xcolor}
\usepackage{caption}
\usepackage{subcaption}
\usepackage{xfrac}
\newcommand{\ccline}[1]{%
    \cline{#1}%
    \noalign{\vskip-\doublerulesep}
    \cline{#1}%
    \noalign{\vskip\doublerulesep}
}
\def\pmtt{\texttt{+\!\!\!\raisebox{-.48ex}-}}

\def\ttlf{\ttfamily\fontseries{l}\selectfont}
\usepackage{lmodern}
\usepackage{ifthen}
\usepackage{tabularx}
\usepackage{colortbl}
\usepackage{booktabs}
\usepackage{pos}
\usepackage{cleveref}

\definecolor{babypink}{rgb}{0.96, 0.76, 0.76}

\newcommand{\ii}{\mathrm{i}}
\PassOptionsToPackage{export}{adjustbox}
\chardef\MyArticleWithColor=\pdfcolorstackinit page direct{0 g}

\def\ltd{\mathcal{I}}

\def\thresholdct{\mathcal{C}}

\def\unitvector{\hat{\mathbf{k}}}
\def\threshold{t}
\def\origin{\vec{\mathbf{O}}}
\def\loopmomenta{\vec{\mathbf{k}}}
\def\thresholdset{T}

\def\centerarc[#1](#2)(#3:#4:#5)
    { \draw[#1] ($(#2)+({#5*cos(#3)},{#5*sin(#3)})$) arc (#3:#4:#5); }
\title{Numerical integration of the double- and triple-box integrals using threshold subtraction}

\author[a,b]{Dario Kermanschah}
\author*[c]{Matilde Vicini}

\affiliation[a]{Physics Department, Technical University Munich, James-Franck-Strasse 1, 85748 Garching, Germany}
\affiliation[b]{Rudolf Peierls Centre for Theoretical Physics, Oxford University, Clarendon Laboratory, Parks Road, Oxford OX1 3PU, UK}
\affiliation[c]{Institute for Theoretical Physics, ETH Zurich, Wolfgang-Pauli-Strasse 27, 8093 Z\"urich, Switzerland}

\emailAdd{d.kermanschah@gmail.com, mvicini@phys.ethz.ch}

\abstract{
We numerically integrate finite two- and three-loop scalar integrals
using the threshold subtraction method~\cite{Kermanschah:2021wbk}. 
This represents a first step towards extending our calculation of the $N_f$-part to the full NNLO virtual corrections for the production of massive vector bosons presented in~\cite{kermanschah2024nfcontributionvirtualcorrectionelectroweak}. 
}

\FullConference{Loops and Legs in Quantum Field Theory (LL2024)\\
 14-19, April, 2024\\
Wittenberg, Germany\\}

\begin{document}
\maketitle
\tikzset{->-/.style={decoration={
  markings,
  mark=at position {#1+0.1} with {\arrow{latex}}},postaction={decorate}}}
\section{Introduction}
The construction of locally finite amplitudes of refs.~\cite{Anastasiou:2020sdt,Anastasiou:2022eym,Anastasiou:2024xvk} calls for an efficient method to compute loop integrals numerically in momentum space. 
A viable approach is based on the subtraction of threshold singularities, which originated in~\cite{Kilian:2009wy} and was later extended in~\cite{Kermanschah:2021wbk} for the intersecting singularity structures.
In conjunction with the local infrared~(IR) and ultraviolet~(UV) subtraction of~\cite{Anastasiou:2022eym}, threshold subtraction has proven to be effective for the computation of $N_f$-contributions to the virtual correction at NNLO for the production of massive electroweak vector bosons at the LHC~\cite{kermanschah2024nfcontributionvirtualcorrectionelectroweak}.
In this contribution, we apply the threshold subtraction method to scalar two- and three-loop integrals: the four-point ladder diagrams with massive external legs and massless internal propagators.
These integrals are finite in IR and UV limits and can be evaluated directly in $d=4$ spacetime dimensions.
Furthermore, these integrals allow us to test the threshold subtraction method  for the first time for two-~and~three-loop type thresholds without introducing the added complexity of multi-channelling, as we will show below.
These calculations pave the way for future applications, such as extending the work in ref.~\cite{kermanschah2024nfcontributionvirtualcorrectionelectroweak} to the complete virtual cross section at NNLO.
\section{Method}
We consider the double-~and~triple-box integrals
\begin{equation}\label{eq:loop-integrals-ladder}
G^{(2)}= \int \frac{\dd^4k_1}{(2\pi)^4}\frac{\dd^4k_2}{(2\pi)^4}
\frac{1}{A_1\cdots A_7}\,, \qquad G^{(2)}= \int \frac{\dd^4k_1}{(2\pi)^4}\frac{\dd^4k_2}{(2\pi)^4}\frac{\dd^4k_3}{(2\pi)^4}
\frac{1}{B_1\cdots B_{10}} \,,
\end{equation}
with 
\begin{alignat}{2}
    A_1&=B_1= k_1^2\,,\qquad
    &&A_2=B_2= (k_1-p_1-p_2)^2\,,\\
    A_3&=B_3= (k_1-p_2)^2\,,\qquad
    &&A_4=B_4= (k_1-k_2)^2\,,\\
    A_5&=B_5= k_2^2\,,\qquad
    &&A_6=B_6= (k_2-p_1-p_2)^2\,,\\
    A_7&=(k_2-q_1)^2\,,\qquad
    &&B_7=(k_2-k_3)^2\,,\qquad
    B_8=k_3^2\,,\\
    B_9&=(k_3-q_1)^2\,,\qquad
    &&B_{10}=(k_3-p_1-p_2)^2\,,
\end{alignat}
where the $\ii\epsilon$ causal prescription is left implicit and 
$p_1+p_2=q_1+q_2$, $p_i^2=m_i^2\neq0$, $q_i^2=m_i^2\neq0$\,.
\par
Despite being IR- and UV-finite, these integrals remain difficult to evaluate using Monte Carlo integration.
The  $\ii \epsilon$ prescription, which formally shifts the on-shell poles of the propagators away from the contour of integration, is impractical for numerical integration.
We summarise how this issue is resolved, following refs.~\cite{Kermanschah:2021wbk,kermanschah2024nfcontributionvirtualcorrectionelectroweak}. In section~\ref{subsec:energy-int}, we analytically integrate the energy component of the loop momenta, in order to expose the threshold singular surfaces. 
In section~\ref{subsec:threshold-sing}, we regularise the corresponding threshold singularities.
\subsection{Integration of loop energies}\label{subsec:energy-int}
Time-ordered perturbation theory (TOPT) \cite{Bodwin:1984hc, Collins:1985ue, Sterman:1993hfp, Sterman:1995fz}, loop-tree duality (LTD)~\cite{Catani:2008xa, Aguilera-Verdugo:2019kbz, JesusAguilera-Verdugo:2020fsn, Ramirez-Uribe:2022sja, Runkel:2019yrs, Capatti:2019ypt}, causal LTD~\cite{Aguilera-Verdugo:2020set, Aguilera-Verdugo:2020kzc, Ramirez-Uribe:2020hes, Sborlini:2021owe, TorresBobadilla:2021ivx, Bobadilla:2021pvr, Benincasa:2021qcb, Kromin:2022txz, Capatti:2020ytd}, the cross-free family (CFF) representation~\cite{Capatti:2022mly, Capatti:2023shz}, and partially time-ordered perturbation theory (PTOPT) \cite{Sterman:2023xdj} all provide a systematic way of performing the energy integration, 
but they yield different algebraic representations (of the same expression). 
Their derivation exploits that each (quadratic) propagator denominator
\begin{equation}\label{eq:props}
    D_i = q_i^2 - m_i^2+
    \ii \epsilon =(q_i^0 - E_i)(q_i^0 + E_i)
\end{equation}
has two poles at positive and negative on-shell energies $q^0_i=\pm E_i$\,.
After the integration over the energy component of the loop momenta, the integrand is given by
\begin{equation}\label{eq:Iltddef}
\int\limits \left(\prod_{j=1}^{n} \frac{\dd k_j^0}{(2\pi)}\right)
    \mathcal{G}^{(n)}
    \left(\{k_i\}\right)
    =  \left( -\ii\right)^n
    \ltd^{(n)}
    \left(\{\vec{k}_i\}\right)\,,
\end{equation}
where $\mathcal{G}^{(n)}$ denotes the integrand of $G^{(n)}$.
$\ltd^{(n)}$ is a rational function of the on-shell energies, whose integral over the spatial loop momentum space we denote as $I^{(n)}$. 
The remaining poles are precisely the threshold singularities that are addressed in the next section.
\par
We will implement our expressions as in ref.~\cite{kermanschah2024nfcontributionvirtualcorrectionelectroweak}.
Since CFF, unlike LTD, has no spurious singularities, it is more numerically stable and it is therefore our preferred representation for numerical evaluation.
On the other hand, the LTD representation is more compact and we use it for the residues needed for the threshold counterterms presented below in eq.~\eqref{eq:ct}.
We generate these expressions using \textsc{Form}~\cite{Vermaseren:2000nd,Kuipers:2013pba,Ruijl:2017dtg,Ueda:2020wqk} and \textsc{Python}, as described in~\cite{kermanschah2024nfcontributionvirtualcorrectionelectroweak}. 
\subsection{Subtraction of threshold singularities}\label{subsec:threshold-sing}
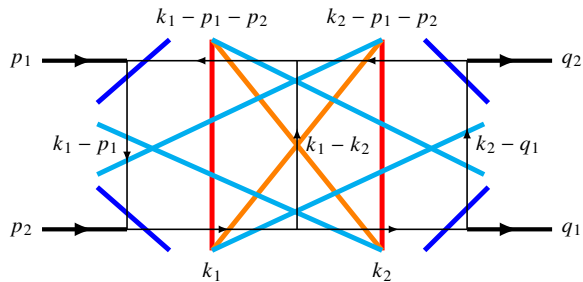
\begin{figure}[b]

    \centering
    \resizebox{!}{3.8cm}{
    \begin{tikzpicture}[scale=3]
        \coordinate (p2a) at (0.5,0);
        \coordinate (p1a) at (0.5,1);
        \coordinate (p2b) at (1,0);
        \coordinate (p1b) at (1,1);
 \coordinate (p1b) at (1,1);

 \coordinate(v5) at (2,0);
 \coordinate(v6) at (2,1);
        
        \coordinate (Q2a) at (3,1);
        \coordinate (Q2b) at (3.5,1);
        \coordinate (Q1a) at (3,0);
         \coordinate (Q1b) at (3.5,0);
           \coordinate (q3a) at (3,0.5);
         \coordinate (q3b) at (3.5,0.5);
      \draw[line width=0.75mm,->-=0.5] (p1a) node[left] {$p_1$}  -- (p1b);
        \draw[line width=0.75mm,->-=0.5] (Q2a) --  (Q2b) node[right] {$q_2$};
        \draw[line width=0.75mm,->-=0.5] (p2a) node[left] {$p_2$}  --  (p2b);
        \draw[line width=0.75mm,->-=0.5] (Q1a) -- (Q1b) node[right] {$q_1$};

        \draw [line width=2.5pt,red,opacity=0.6] (1.5,1.125) -- (1.5,-0.125);
         \draw [line width=2.5pt,red,opacity=0.6] (2.5,1.125) -- (2.5,-0.125);
         \draw [line width=2.5pt,orange,opacity=0.6] (2.5,1.125) -- (1.5,-0.125);
         \draw [line width=2.5pt,orange,opacity=0.6] (1.5,1.125) -- (2.5,-0.125);

        \draw [line width=2.5pt,blue,opacity=0.6] (2.75,1.125) -- (3.125,0.75);
        \draw [line width=2.5pt,cyan,opacity=0.6] (1.5,1.125) --  (3.1,0.325);
        \draw [line width=2.5pt,blue,opacity=0.6] (2.75,-0.125) -- (3.125,0.25);
        \draw [line width=2.5pt,cyan,opacity=0.6] (1.5,-0.125) --  (3.1,0.625);

        \draw [line width=2.5pt,cyan,opacity=0.6] (0.825,0.625) --  (2.5,-0.125);
       \draw [line width=2.5pt,blue,opacity=0.6] (1.25,-0.125) --  (0.825,0.25);
                \draw [line width=2.5pt,cyan,opacity=0.6] (0.825,0.325) --  (2.5,1.125);
                \draw [line width=2.5pt,blue,opacity=0.6] (1.25,1.125) --  (0.825,0.75);
    \draw[thick,->-=0.5] (p1b) -- (p2b) node[midway,left] {$k_1-p_1$};
\draw[thick,->-=0.5] (Q1a) -- (Q2a) node[midway,right] {$k_2-q_1$};
\draw[thick,->-=0.5] (v6) -- (p1b) node[midway,above=0.45cm] {$k_1-p_1-p_2$};
\draw[thick,->-=0.5] (p2b) -- (v5) node[midway,below=0.45cm] {$k_1$};
    \draw[thick,->-=0.5] (v5) -- (Q1a) node[midway,below=0.45cm] {$k_2$};
    \draw[thick,->-=0.5] (Q2a) -- (v6) node[midway,above=0.45cm] {$k_2-p_1-p_2$};
         \draw[thick,->-=0.5] (v5) -- (v6) node[midway,right] {$k_1-k_2$};
  \end{tikzpicture}
  }
  \caption{Cutkosky cuts of the double-box diagram with massive external legs.
  }
  \label{fig:ladder2}
  \end{figure}

\begin{figure}[b]
    \centering
    \resizebox{!}{3.8cm}{
    \begin{tikzpicture}[scale=3]
        \coordinate (p2a) at (0.5,0);
        \coordinate (p1a) at (0.5,1);
        \coordinate (p2b) at (1,0);
        \coordinate (p1b) at (1,1);
 \coordinate (p1b) at (1,1);

 \coordinate(v5) at (2,0);
 \coordinate(v6) at (2,1);
  \coordinate(v7) at (3,0);
 \coordinate(v8) at (3,1);    
 
        \coordinate (Q2a) at (4,1);
        \coordinate (Q2b) at (4.5,1);
        \coordinate (Q1a) at (4,0);
         \coordinate (Q1b) at (4.5,0);
           \coordinate (q3a) at (4,0.5);
         \coordinate (q3b) at (4.5,0.5);
      \draw[line width=0.75mm,->-=0.5] (p1a) node[left] {$p_1$}  -- (p1b);
        \draw[line width=0.75mm,->-=0.5] (Q2a) --  (Q2b) node[right] {$q_2$};
        \draw[line width=0.75mm,->-=0.5] (p2a) node[left] {$p_2$}  --  (p2b);
        \draw[line width=0.75mm,->-=0.5] (Q1a) -- (Q1b) node[right] {$q_1$};
        
        \draw [line width=2.5pt,red,opacity=0.6] (1.5,1.125) -- (1.5,-0.125);
         \draw [line width=2.5pt,red,opacity=0.6] (2.5,1.125) -- (2.5,-0.125);
         \draw [line width=2.5pt,red,opacity=0.6] (3.5,1.125) -- (3.5,-0.125);
        \draw [line width=2.5pt,orange,opacity=0.6] (1.5,1.125) -- (2.5,-0.125);
        \draw [line width=2.5pt,pink,opacity=0.6] (1.5,1.125) -- (3.5,-0.125);
 \draw [line width=2.5pt,orange,opacity=0.6] (2.5,1.125) -- (1.5,-0.125);
          \draw [line width=2.5pt,pink,opacity=0.6] (3.5,1.125) -- (1.5,-0.125);
\draw [line width=2.5pt,orange,opacity=0.6] (2.5,1.125) -- (3.5,-0.125);
        \draw [line width=2.5pt,orange,opacity=0.6] (3.5,1.125) -- (2.5,-0.125);  
        \draw [line width=2.5pt,blue,opacity=0.6] (3.75,1.125) -- (4.125,0.75);
        \draw [line width=2.5pt,cyan,opacity=0.6] (2.5,1.125) --  (4.1,0.325);
        \draw [line width=2.5pt,green,opacity=0.6] (1.25,1.125) --  (4.1,0.325);
        \draw [line width=2.5pt,blue,opacity=0.6] (3.75,-0.125) -- (4.125,0.25);
        \draw [line width=2.5pt,cyan,opacity=0.6] (2.5,-0.125) --  (4.1,0.625);
\draw [line width=2.5pt,green,opacity=0.6] (1.25,-0.125) --  (4.1,0.625);
        \draw [line width=2.5pt,cyan,opacity=0.6] (0.825,0.625) --  (2.5,-0.125);
        \draw [line width=2.5pt,green,opacity=0.6] (0.825,0.625) --  (3.725,-0.125);
       \draw [line width=2.5pt,blue,opacity=0.6] (1.25,-0.125) --  (0.825,0.25);
                \draw [line width=2.5pt,cyan,opacity=0.6] (0.825,0.325) --  (2.5,1.125);
                \draw [line width=2.5pt,green,opacity=0.6] (0.825,0.325) --  (3.725,1.125);
                \draw [line width=2.5pt,blue,opacity=0.6] (1.25,1.125) --  (0.825,0.75);
  \draw[thick,->-=0.5] (p1b) -- (p2b) node[midway,left] {$k_1-p_1$};
\draw[thick,->-=0.5] (v6) -- (p1b) node[midway,above=0.45cm] {$k_1-p_1-p_2$};
\draw[thick,->-=0.5] (p2b) -- (v5) node[midway,below=0.45cm] {$k_1$};
    \draw[thick,->-=0.5] (v5) -- (v7) node[midway,below=0.45cm] {$k_2$};
    \draw[thick,->-=0.5] (v8) -- (v6) node[midway,above=0.45cm] {$k_2-p_1-p_2$};
    \draw[thick,->-=0.5] (v5) -- (v6) node[midway,right] {$k_1-k_2$};
    \draw[thick,->-=0.5] (v7) -- (v8) node[midway,right] {$k_2-k_3$};
     \draw[thick,->-=0.5] (v7) -- (Q1a) node[midway,below=0.45cm] {$k_3$};
     \draw[thick,->-=0.5] (Q1a) -- (Q2a) node[midway,right] {$k_3-q_1$};
     \draw[thick,->-=0.5] (Q2a) -- (v8) node[midway,above=0.45cm] {$k_3-p_1-p_2$};
  \end{tikzpicture}
  }
  \caption{Cutkosky cuts of the triple-box diagram with massive external legs.}
    \label{fig:ladder3}
\end{figure}
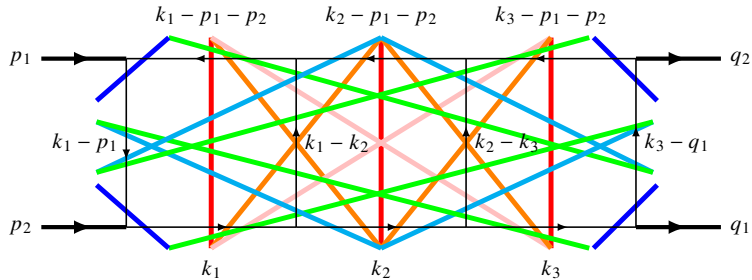

For our scalar integrals, we find the threshold singularities listed as Cutkosky cuts~\cite{Cutkosky:1960sp} in figs.~\ref{fig:ladder2}~and~\ref{fig:ladder3}. Alternatively, they can be determined from the denominator structure in the CFF or LTD expression.
The threshold singularities illustrated in fig.~\ref{fig:ladder2} are 
\begin{alignat}{2}
\label{eq:t-start}
	\textcolor{red}{\threshold_1} &=  E_2+E_1-p^0_2-p^0_1            \,, \qquad 
&& \textcolor{red}{\threshold_2} =  E_6+E_5-p^0_2-p^0_1            \,,\\ 
 	\textcolor{orange}{\threshold_3} &=  E_2+E_4+E_5-p^0_2-p^0_1        \,,\qquad
&& \textcolor{orange}{\threshold_4} =  E_6+E_4+E_1-p^0_2-p^0_1        \,, \\ 
 \label{eq:t5-6}
	\textcolor{blue}{\threshold_5} &=  E_2+E_3-p^0_1                  \,,\qquad
	&&\textcolor{cyan}{\threshold_6} =  E_6+E_4+E_3-p^0_1             \,,\\
		\textcolor{blue}{\threshold_7}  &=  E_3+E_1-p^0_2 \,,\qquad
 &&\textcolor{cyan}{\threshold_8} =  E_4+E_3+E_5-p^0_2             \,,\\
 \textcolor{blue}{\threshold_9} &=  E_7+E_5-q^0_1 \,,\qquad
	 &&\textcolor{cyan}{\threshold_{10}} =  E_7+E_4+E_1-q^0_1              \,,\\
 	 \textcolor{blue}{\threshold_{11}}&=  E_6+E_7+q^0_1-p^0_2-p^0_1      \,,\qquad   
	 &&\textcolor{cyan}{\threshold_{12}}=  E_2+E_7+E_4+q^0_1-p^0_2-p^0_1  \,,
  \label{eq:t-end}
\end{alignat}
where
\begin{align*}
E_1&=|\vec k_1| \,,\quad E_5=|\vec k_2|\,,\quad E_3=|\vec k_1-\vec p_1|\,,\quad E_4=|\vec k_1-\vec k_2|
\,,\\
E_2&=|\vec k_1-\vec p_1-\vec p_2|
\,,\quad E_6=|\vec k_2-\vec p_1-\vec p_2|
\,,\quad E_7=|\vec k_2-\vec q_1|\,.
\end{align*}
Analogously for $\ltd^{(3)}$, we find the twenty-one threshold singularities represented in fig.~\ref{fig:ladder3}, which we omit listing explicitly for brevity.
\par
We refer to a threshold that contains a sum of $n+1$ on-shell energies $E_i$ as an $n$-loop type threshold. 
The real solutions of $\threshold=0$, where $\threshold$ is one of the above threshold singularities, correspond to poles of the integrand inside the integration domain.
They exist because $p^2>0$, where $p$ is the momentum flowing through the cut.
If $p^2=0$\,, they would describe a pinch singularity, which we do not have in our examples.
\par
Following refs.~\cite{Kermanschah:2021wbk,kermanschah2024nfcontributionvirtualcorrectionelectroweak}, we will parameterise the threshold singularities $\threshold$ as $r=r_\threshold(\unitvector)$, where $r$ is the radial variable of hyperspherical coordinates $(\vec{k}_1,\dots,\vec{k}_n)=\loopmomenta=r \unitvector$. We will remove these poles at $r=r_\threshold(\unitvector)$ using local counterterms.
This approach allows us to determine the dispersive and absorptive parts of $G^{(n)}$ through separate Monte Carlo integrations.
More details on the hypercube mapping in the Monte Carlo integration can be found in ref.~\cite{kermanschah2024nfcontributionvirtualcorrectionelectroweak}.
\par
We can express the dispersive part of $I^{(n)}$ as an integral over $n$ spatial loop momenta,
\begin{align}
\label{eq:dispersive}
    \Re I^{(n)} &= 
    \int \frac{\dd[3n]{\loopmomenta}}{(2\pi)^{3n}}
    \left(
    \ltd^{(n)}(\loopmomenta)
    - \sum_{\threshold \in\thresholdset} \thresholdct_\threshold[\ltd^{(n)}](\loopmomenta)
    \right)\,,
\end{align}
and the absorptive part as 
\begin{align}
    \label{eq:absorptive}
    2\Im I^{(n)} &=
    \int \frac{\dd[3n-1]{\unitvector}}{(2\pi)^{3n-1}}
    \sum_{\threshold \in \thresholdset} 
    R_t[\ltd^{(n)}](\unitvector)\,,
\end{align}
where $\thresholdset$ denotes the set of all threshold singularities.
The threshold counterterm $\thresholdct_t$ is defined as
\begin{align}\label{eq:ct}
    \thresholdct_\threshold
    \left[
    \ltd^{(n)}(r\unitvector)
    \right]
    &=
    \frac{R_\threshold[\ltd^{(n)}]}{r^{3n-1}}
    \left(
    \frac{
        \chi(r-r_\threshold)
    }{
        r-r_\threshold -\ii\epsilon
        }
    +
    \frac{
    \chi(-r-r_\threshold)
    }{
        -r-r_\threshold -\ii\epsilon
        }
    \right),
    \qquad
\end{align}
where we choose
\begin{align}
    \chi(x) = \exp\left[-\left(\frac{x}{E_\text{CM}}\right)^2\right]
    \Theta\left[(3 E_\text{CM})^2-x^2\right]\,,
\end{align}
although any function $\chi$ would suffice to regulate the UV behaviour of $\thresholdct_\threshold$, provided that $\chi$ satisfies $\chi(0)=1$ and $\chi(x)\to 0$ for $x\to \pm\infty$ and is symmetric around the origin.
With these requirements, the integrated counterterm only contributes to the absorptive part of the integral since the Cauchy principal value vanishes by construction, i.e.
\begin{align}
   \int_0^\infty \dd{r} r^{3n-1} \thresholdct_\threshold\left[\ltd^{(n)}(r\unitvector)\right] = \ii \pi R_\threshold[\ltd^{(n)}](\unitvector)\,.
\end{align}
The residue $R_t$ reads
\begin{align}
\label{eq:ps_residue}
    R_\threshold[\ltd^{(n)}(\loopmomenta)] = \Res[r^{3n-1}\ltd^{(n)}(r\unitvector),r=r_\threshold(\unitvector)]
    &= 
    \left. 
    \frac{r^{3n-1}}{\frac{\dd \threshold}{\dd r}}
    \lim_{\threshold\to 0} \left(\threshold  \ltd^{(n)}\right)
    \right\vert_{r=r_\threshold(\unitvector)}\,,
\end{align}
where the threshold parametrisation $r_\threshold(\unitvector)$ is defined implicitly through the on-shell condition
\begin{align}
\label{eq:threshold_param}
\threshold(r_\threshold(\unitvector)\unitvector) = 0\,, \qquad \textrm{where} \qquad r_\threshold(\unitvector)>0\,.
\end{align}
$r_\threshold(\unitvector)$ is the location of the threshold singularity along the direction $\unitvector$.
For the one-loop type thresholds, the on-shell condition of eq.~\eqref{eq:threshold_param} can be solved analytically in the radial variable as explained in ref.~\cite{Kermanschah:2021wbk}.
In general, the solution to the on-shell condition of eq.~\eqref{eq:threshold_param} for higher-loop type thresholds can always be found numerically, for example using Newton's or Brent's method~\cite{Brent1971AnAW}.
However, in our examples and in the centre-of-mass (COM) frame, all threshold singularities collapse to a quadratic equation in $r$, which can easily be solved analytically at fixed $\unitvector$.
\par 
Each threshold singularity describes the locus of points such that $\threshold(\loopmomenta)=0$, hence it defines a hyper-surface in (spatial) loop momentum space.
The above equations are only guaranteed to be correct and the $\ii \epsilon$ causal prescription can be safely removed if the origin $\origin$ of the spherical coordinate system lies inside all threshold surfaces, i.e.  $t(\origin)<0$ for all $\threshold \in \thresholdset$\,. 
Otherwise, the resulting parametrisation would introduce integrable singularities at the tangents of the threshold surfaces, where $\frac{\dd\threshold}{\dd r}(\unitvector)=0$, in the residues of eq.~\eqref{eq:ps_residue}. 
More importantly, if the origin lies inside all the threshold surfaces, it is assured that the counterterms are also correct where the integrand develops higher-order poles, namely at intersections of thresholds surfaces. 
At these intersections, the corresponding residues will also have poles, which, however, locally cancel among each other, as shown in ref.~\cite{Kermanschah:2021wbk}.
\par
If no point inside all intersecting threshold surfaces can be found, one can employ a \textit{multi-channelling} procedure, as in ref.~\cite{kermanschah2024nfcontributionvirtualcorrectionelectroweak}.
Fortunately, in our examples, the point $\vec{0}$ lies inside all threshold surfaces simultaneously. We can therefore use the above equations directly without the need for multi-channelling for the double- and triple-box in the COM frame, by setting $\origin=\vec 0$.
\par
We remark that eq.~\eqref{eq:absorptive} is a local manifestation of the Cutkosky rule~\cite{Cutkosky:1960sp}, in the same way as was shown for its one-loop analogue in ref.~\cite{Kermanschah:2021wbk}.
\section{Results}
Our results are presented in table~\ref{tab:0}.
We use the Vegas adaptive Monte-Carlo algorithm  \cite{Lepage:1977sw,Lepage:2020tgj} of the \textsc{Cuba} library \cite{Hahn:2004fe} for multidimensional numerical integration. 
The solution to the on-shell condition, $r_\threshold(\unitvector)$\,, upon which the threshold counterterms (and residues) depend, is found analytically for all one-loop type thresholds~\cite{Kermanschah:2021wbk}.  The remaining on-shell conditions we solve numerically using Brent's method implemented in \textsc{Rust}~\cite{roots-rust}, since we have not yet implemented higher-loop type analytic solutions that are valid only in specific reference frames.
However, with the analytic solution, we expect an improvement in runtime and stability.
\begin{table}[H]
\centering
\resizebox{\columnwidth}{!}{%
\begin{tabular}{lllrrrccr}
\hline\hline
Diag & Kin & Phase & $N_p \ [10^6]$ & $\sfrac{t}{p} \ [\mu \text{s}]$ & Exp. & Reference  & Result  & $\Delta \ [\%]$ \\
\hline\hline\multirow{6}{*}{\texttt{2L4P}}
& \multirow{2}{*}{\texttt{K1}}
& \texttt{Re}
& {\ttlf\texttt{108}}
& {\ttlf\texttt{0.3}}
& {\ttlf$\texttt{10}^{\texttt{-6}}$}
& {\ttlf\texttt{-1.0841}}
& {\ttlf\texttt{-1.0829~$\pmtt$~0.0054}}
& {\ttlf\texttt{0.495}}
\\
\cline{3-3}\cline{4-4}\cline{5-5}\cline{6-6}\cline{7-7}\cline{8-8}\cline{9-9}& & \texttt{Im}
& {\ttlf\texttt{11}}
& {\ttlf\texttt{0.1}}
& {\ttlf$\texttt{10}^{\texttt{-6}}$}
& {\ttlf\texttt{ 2.8682}}
& {\ttlf\texttt{~2.8651~$\pmtt$~0.0071}}
& {\ttlf\texttt{0.249}}
\\
[\doublerulesep]\ccline{2-2}\ccline{3-3}\ccline{4-4}\ccline{5-5}\ccline{6-6}\ccline{7-7}\ccline{8-8}\ccline{9-9}& \multirow{2}{*}{\texttt{K2}}
& \texttt{Re}
& {\ttlf\texttt{1083}}
& {\ttlf\texttt{0.3}}
& {\ttlf$\texttt{10}^{\texttt{-8}}$}
& {\ttlf\texttt{ 3.1105}}
& {\ttlf\texttt{~3.1091~$\pmtt$~0.0154}}
& {\ttlf\texttt{0.495}}
\\
\cline{3-3}\cline{4-4}\cline{5-5}\cline{6-6}\cline{7-7}\cline{8-8}\cline{9-9}& & \texttt{Im}
& {\ttlf\texttt{14}}
& {\ttlf\texttt{0.1}}
& {\ttlf$\texttt{10}^{\texttt{-8}}$}
& {\ttlf\texttt{ 9.5389}}
& {\ttlf\texttt{~9.5746~$\pmtt$~0.0422}}
& {\ttlf\texttt{0.441}}
\\
[\doublerulesep]\ccline{2-2}\ccline{3-3}\ccline{4-4}\ccline{5-5}\ccline{6-6}\ccline{7-7}\ccline{8-8}\ccline{9-9}& \multirow{2}{*}{\texttt{K3}}
& \texttt{Re}
& {\ttlf\texttt{746}}
& {\ttlf\texttt{0.3}}
& {\ttlf$\texttt{10}^{\texttt{-10}}$}
& {\ttlf\texttt{ 1.7037}}
& {\ttlf\texttt{~1.7142~$\pmtt$~0.0085}}
& {\ttlf\texttt{0.496}}
\\
\cline{3-3}\cline{4-4}\cline{5-5}\cline{6-6}\cline{7-7}\cline{8-8}\cline{9-9}& & \texttt{Im}
& {\ttlf\texttt{13}}
& {\ttlf\texttt{0.1}}
& {\ttlf$\texttt{10}^{\texttt{-10}}$}
& {\ttlf\texttt{ 4.5650}}
& {\ttlf\texttt{~4.5620~$\pmtt$~0.0210}}
& {\ttlf\texttt{0.461}}
\\
\hline\hline\multirow{6}{*}{\texttt{3L4P}}
& \multirow{2}{*}{\texttt{K1}}
& \texttt{Re}
& {\ttlf\texttt{982}}
& {\ttlf\texttt{4\phantom{.0}}}
& {\ttlf$\texttt{10}^{\texttt{-9}}$}
& {\ttlf\texttt{-2.4242}}
& {\ttlf\texttt{-2.4042~$\pmtt$~0.0204}}
& {\ttlf\texttt{0.849}}
\\
\cline{3-3}\cline{4-4}\cline{5-5}\cline{6-6}\cline{7-7}\cline{8-8}\cline{9-9}& & \texttt{Im}
& {\ttlf\texttt{10008}}
& {\ttlf\texttt{6\phantom{.0}}}
& {\ttlf$\texttt{10}^{\texttt{-9}}$}
& {\ttlf\texttt{-3.4003}}
& {\ttlf\texttt{-3.4037~$\pmtt$~0.0298}}
& {\ttlf\texttt{0.874}}
\\
[\doublerulesep]\ccline{2-2}\ccline{3-3}\ccline{4-4}\ccline{5-5}\ccline{6-6}\ccline{7-7}\ccline{8-8}\ccline{9-9}& \multirow{2}{*}{\texttt{K2}}

& \texttt{Re}
& {\ttlf\texttt{3763}}
& {\ttlf\texttt{5\phantom{.0}}}
& {\ttlf$\texttt{10}^{\texttt{-11}}$}
& {\ttlf\texttt{-5.3031}}
& {\ttlf\texttt{-5.2649~$\pmtt$~0.0447}}
& {\ttlf\texttt{0.848}}
\\
\cline{3-3}\cline{4-4}\cline{5-5}\cline{6-6}\cline{7-7}\cline{8-8}\cline{9-9}& & \texttt{Im}
& {\ttlf\texttt{10008}}
& {\ttlf\texttt{6\phantom{.0}}}
& {\ttlf$\texttt{10}^{\texttt{-11}}$}
& {\ttlf\texttt{-1.0780}}
& {\ttlf\texttt{-1.1501~$\pmtt$~0.1433}}
& {\ttlf\texttt{{12.459}}}
\\
[\doublerulesep]\ccline{2-2}\ccline{3-3}\ccline{4-4}\ccline{5-5}\ccline{6-6}\ccline{7-7}\ccline{8-8}\ccline{9-9}& \multirow{2}{*}{\texttt{K3}}
& \texttt{Re}
& {\ttlf\texttt{4303}}
& {\ttlf\texttt{7\phantom{.0}}}
& {\ttlf$\texttt{10}^{\texttt{-14}}$}
& {\ttlf\texttt{-4.4705}}
& {\ttlf\texttt{-4.4283~$\pmtt$~0.0376}}
& {\ttlf\texttt{0.849}}
\\
\cline{3-3}\cline{4-4}\cline{5-5}\cline{6-6}\cline{7-7}\cline{8-8}\cline{9-9}& & \texttt{Im}
& {\ttlf\texttt{10008}}
& {\ttlf\texttt{5\phantom{.0}}}
& {\ttlf$\texttt{10}^{\texttt{-15}}$}
& {\ttlf\texttt{-6.6383}}
& {\ttlf\texttt{-6.3589~$\pmtt$~1.3079}}
& {\ttlf\texttt{{20.568}}}
\\
\cline{3-3}\cline{4-4}\cline{5-5}\cline{6-6}\cline{7-7}\cline{8-8}\cline{9-9}\hline\hline
\end{tabular}%
}%
\caption{\label{tab:0}
Results for numerical integration of double- and triple-box integrals, respectively denoted as \texttt{2L4P}, \texttt{3L4P}, at the respective kinematical points listed in table~\ref{tab:kinematics}\,, with reference results from ref.~\cite{Usyukina:1992jd}.
$N_p$ denotes the number of Monte Carlo evaluations, $\sfrac{t}{p}$~represents the average time per evaluation, $\text{Exp.}$ indicates the scale by which the result is multiplied, and $\Delta$ signifies the relative uncertainty of the result.
The Vegas parameters \texttt{nstart} and \texttt{nincrease} are set to $\texttt{10}^{\texttt{6}}$ and $\texttt{10}^{\texttt{5}}$ respectively,
\texttt{nmax} is set to $\texttt{10}^{\texttt{10}}$, \texttt{epsrel} to $\texttt{0.085}$\,.
The numerical integrations were performed on a standard computer with an AMD Ryzen 9 5950X 16-Core Processor CPU, on 15 cores.
}%
\end{table}

The reference results in table~\ref{tab:0} are from the analytic computation of ref.~\cite{Usyukina:1992jd}. 
The example kinematic configurations \texttt{K1}, \texttt{K2}, and \texttt{K3} are taken from~\cite{Capatti:2019edf}, where these integrals were evaluated using contour deformation. 
For completeness, they are listed in table~\ref{tab:kinematics}.
The three-loop ladder diagram's dispersive part evaluation requires the rescue of unstable samples in quadruple precision. 
For this diagram, we checked the stability of the integrand evaluation using a rotation test, as described in \cite{Capatti:2019edf}. This test involves comparing the integrand evaluation at the original kinematic configuration and at a rotated one. If a specified minimum number of digits match, the evaluation is deemed stable. Additionally, the rotation test can be repeated in quadruple precision to potentially recover an unstable sample. If the evaluation remains unstable, it is set to zero. The number of unstable and rescued samples is tracked, and the count of unstable samples should not exceed a certain bound to avoid distorting the Monte Carlo estimate.
It remains to be investigated why the errors are large in the dispersive integral of \texttt{3L4P} for kinematic configurations \texttt{K2} and \texttt{K3}\,, a curiosity that was already observed in ref.~\cite{Capatti:2019edf}. It  may be attributed to their larger scale hierarchy with respect to \texttt{K1}. Potentially the method can be adjusted to tackle these kinematic points more efficiently.
\begin{table}[H]
    \centering
    \resizebox{0.95\textwidth}{!}{
    \begin{tabular}{ll}
     \hline\hline
     \noalign{\vskip 0.8mm}
    \texttt{K1} & 
    {  
    $\!\begin{aligned} 
        p_1^{\mu}&=\texttt{(2.50925, 0.0, 0.0, 2.30138)}\\
        p_2^{\mu}&=\texttt{(2.55075, 0.0, 0.0, -2.30138)}\\
        q_1^{\mu}&=\texttt{(2.5053, 0.487891, 1.95655, -0.877716)}
    \end{aligned}$
    }
    \vspace{0.8mm}
    \\
    \hline\hline
    \noalign{\vskip 0.8mm}
    \texttt{K2} & 
    {$\!\begin{aligned} 
        p_1^{\mu}&=\texttt{(6.0, 0.0, 0.0, 5.91607978309962)}\\
        p_2^{\mu}&=\texttt{(6.0, 0.0, 0.0, -5.91607978309962)}\\
        q_1^{\mu}&=\texttt{(6.0, 1.3124738333059, 5.26330888118183, -2.36114210884473)}
    \end{aligned}$}
      \vspace{0.8mm}
    \\ \hline\hline
    \noalign{\vskip 0.8mm}
 \texttt{K3} &  { $\!\begin{aligned} 
        p_1^{\mu}&=\texttt{(14.95, 0.0, 0.0, 14.9165176901313)}\\
        p_2^{\mu}&=\texttt{(15.05, 0.0, 0.0, -14.9165176901313)}\\
        q_1^{\mu}&=\texttt{(14.8833333333333,
                  3.23407440276709,
              12.9693500125724,
                   -5.81810399699641)}
    \end{aligned}$ }
      \vspace{0.8mm}  \\  \hline \hline
    \end{tabular}
    }
    \caption{\label{tab:kinematics} 
    Kinematic points  \texttt{K1}, \texttt{K2}, \texttt{K3} used for the results in table~\ref{tab:0}. From momentum conservation
    $q_2=p_1+p_2-q_1$.}
\end{table}

\section{Conclusion}
We demonstrated the threshold subtraction method in difficult two- and three-loop finite scalar integrals, where a large number of thresholds are present.
In the future, for the extension of ref.~\cite{kermanschah2024nfcontributionvirtualcorrectionelectroweak} to the full NNLO virtual cross section, we will encounter similar two-loop diagrams. 
However, since these diagrams will have massless initial particles, 
some threshold singularities will be traded for pinch singularities, which will instead be subtracted by the local infrared counterterms of ref.~\cite{Anastasiou:2022eym}.
\acknowledgments
M.V. thanks the organisers of LL2024. 
This work was supported by the Swiss National Science Foundation through its Postdoc.Mobility funding scheme (grant number 211062) and project funding scheme (grant number 10001706).
\bibliographystyle{JHEP}
\bibliography{skeleton.bib}

\end{document}